\documentclass[twocolumn,aps,showpacs,floatfix,superscriptaddress]{revtex4}
\pdfoutput=1 
\usepackage{amsmath,amssymb,eucal,graphicx}
 
\begin{document}
\title{Limiting Shapes in Two-Dimensional Ising Ferromagnets}

\author{P.~L.~Krapivsky}
\affiliation{Department of Physics, Boston University, Boston, MA 02215, USA}
\author{Jason Olejarz}
\affiliation{Department of Physics, Boston University, Boston, MA 02215, USA}

\begin{abstract}
We consider an Ising model on a square lattice with ferromagnetic spin-spin interactions spanning beyond nearest neighbors. Starting from initial states with a single unbounded interface separating ordered phases,  we investigate the evolution of the interface subject to zero-temperature spin-flip dynamics. We consider an interface which is initially (i) the boundary of the quadrant, or (ii) the boundary of a semi-infinite stripe. In the former case the interface recedes from its original location in a self-similar diffusive manner. After a re-scaling by $\sqrt{t}$, the shape of the interface becomes more and more deterministic; we determine this limiting shape analytically and verify our predictions numerically. The semi-infinite stripe acquires a stationary shape resembling a finger, and this finger translates along its axis. We compute the limiting shape and the velocity of the Ising finger. 
\end{abstract}  
  

\maketitle

\section{Introduction}

The concept of universality plays a crucial role in understanding of myriads of many-particle systems. Universality is the claim that details are irrelevant, so that an interesting behavior of a given system can be extracted from its simplified versions. In the context of Ising ferromagnets, for instance, universality implies that the critical behavior depends on the spatial dimensionality, but not on the details of the underlying lattice. Universality is also embodied in another even more `obvious' claim, viz. that the toy Ising model with nearest-neighbor (NN) interactions faithfully represents the behavior of the general model with ferromagnetic interactions spanning beyond nearest neighbors. Here we probe the validity of this latter assertion. 

Equilibrium behaviors at the critical temperature agree with universality when the interaction strength quickly decreases with separation between the spins \cite{long}. Dynamical behaviors are critical \cite{LAC,Bray94,book}, as it is manifested e.g. by algebraic time dependencies both at the critical temperature and at sub-critical temperatures. Here we consider the latter, more specifically a zero-temperature spin-flip dynamics. We focus on the two-dimensional ferromagnetic Ising model. On the square lattice, the smallest perturbation of the Ising model with NN interactions accounts for next-nearest-neighbor (NNN) interactions. The Hamiltonian is
\begin{equation}
\label{Ising_H}
\mathcal{H} = - J_1\sum_{\text{NN}} s_{\bf i} s_{\bf j}
- J_2\sum_{\text{NNN diag}} s_{\bf i} s_{\bf j}
\end{equation}
where $s=\pm 1$ are Ising spins. Couplings in Eq.~\eqref{Ising_H} are presumed to be ferromagnetic, $J_1>0$ and $J_2>0$, but otherwise arbitrary. The first sum in Eq.~\eqref{Ising_H} is taken over NN pairs, that is, over sites ${\bf i}=(i_1,i_2)$ and ${\bf j}=(j_1,j_2)$ such that ${\bf i} -{\bf j}$ is equal to $(\pm 1, 0)$ or $(0, \pm 1)$. The second sum in the Hamiltonian \eqref{Ising_H} is taken over NNN pairs of spins, more precisely over diagonal neighbors, that is, ${\bf i} -{\bf j}$ attains one of the four values: $(1, 1),~ (1, -1), ~ (-1, 1),~ (-1, -1)$. 

At zero temperature, energy raising spin flips are never performed. A zero-temperature spin-flip dynamics implied by the Hamiltonian \eqref{Ising_H} quantitatively differs from the standard dynamics. We are interested in qualitative changes, however. Surprisingly, such changes do exist, although they are rather subtle. For instance, starting from above the critical temperature and suddenly quenching to zero temperature, the standard Ising model on the square lattice can get trapped in a state with vertical or horizontal stripes \cite{SKR,ONSS06,BKR09}. For the Ising model characterized by the Hamiltonian \eqref{Ising_H}, stripes going in the $(1,1)$ and $(1,-1)$ directions (the boundaries of such stripes are perfect ladders with steps of the size of the lattice spacings) are also possible \cite{OKR:2d}. Increasing the span of interactions results in positive (in the thermodynamic limit) probabilities of observing more exotic stripe states \cite{OKR:2d}. 

In this work we focus on simple deterministic initial settings, specifically on the behavior of a single interface separating ordered phases. One example is the evolution  of a quadrant of minority phase (Fig.~\ref{def-new}). Both for the standard Ising model and for the model with Hamiltonian \eqref{Ising_H} the major features of the dynamics are the same, e.g. the typical size of the `melted' region grows diffusively, i.e., as $\sqrt{t}$. After re-scaling by $\sqrt{t}$, the boundary of the melted region becomes more and more deterministic and it approaches a limiting shape in the large time limit. 

We will show that the limiting shape arising for the Ising model with Hamiltonian \eqref{Ising_H} differs from the limiting shape corresponding to the standard Ising model. Increasing the interaction span leads to a zoo of limiting shapes. To define the interaction range we endow the lattice (the square lattice in our case) with a metric and postulate that spins separated by distances exceeding the interaction range don't interact. The Manhattan metric $|{\bf i}-{\bf j}|= |i_1 - j_1| + |i_2-j_2|$ is perhaps the simplest. We shall use it and we consider a class of Hamiltonians 
\begin{equation}
\label{Ising_Ham}
\mathcal{H}_k = - \sum_{n=1}^k \sum_{|{\bf i}-{\bf j}|=n} J_n s_{\bf i} s_{\bf j}
\end{equation}
Coupling constants are always assumed to be ferromagnetic, $J_n>0$ for all $n=1,\ldots,k$, and rapidly decreasing with $n$. Once such requirements [\eqref{JJ} and \eqref{JJJ} below] are obeyed, the precise values of the coupling constants are irrelevant and only the interaction range $k$ plays a role. 

The rest of this paper is organized as follows. We study in detail the Ising model which, in addition to the standard NN interactions, has NNN interactions. It is easier to think about the Hamiltonian \eqref{Ising_H} than the Hamiltonian \eqref{Ising_Ham} with $k=2$, although  there is no difference, e.g. in the case of a single interface shown in Fig.~\ref{def-new} both Hamiltonians lead to the same dynamics. In Sec.~\ref{sec:map}, in a specific example of the `melting' of the quadrant of minority phase, we map the interface onto a lattice gas.  We utilize this mapping in Sec.~\ref{sec:quadrant}: We first examine the evolution of the density of the lattice gas, and then translate this knowledge into a calculation of the limiting shape of the interface. In Sec.~\ref{sec:quadrant_field}, we study how an external magnetic field, which disfavors the minority phase occupying the quadrant, changes the limiting shape. The limiting shape of the finger is described in Sec.~\ref{sec:finger}. All the limiting shapes in sections \ref{sec:quadrant}--\ref{sec:finger} are derived in the realm of the Ising ferromagnet with NNN interactions. In Sec.~\ref{sec:range}, we analyze larger interaction ranges, viz. we consider a class of Hamiltonians \eqref{Ising_Ham} with $k\geq 3$. Concluding remarks are presented in Sec.~\ref{concl}.

\section{Mapping of an interface onto a lattice gas}
\label{sec:map}

\begin{figure}
\centering
\includegraphics[scale=0.33]{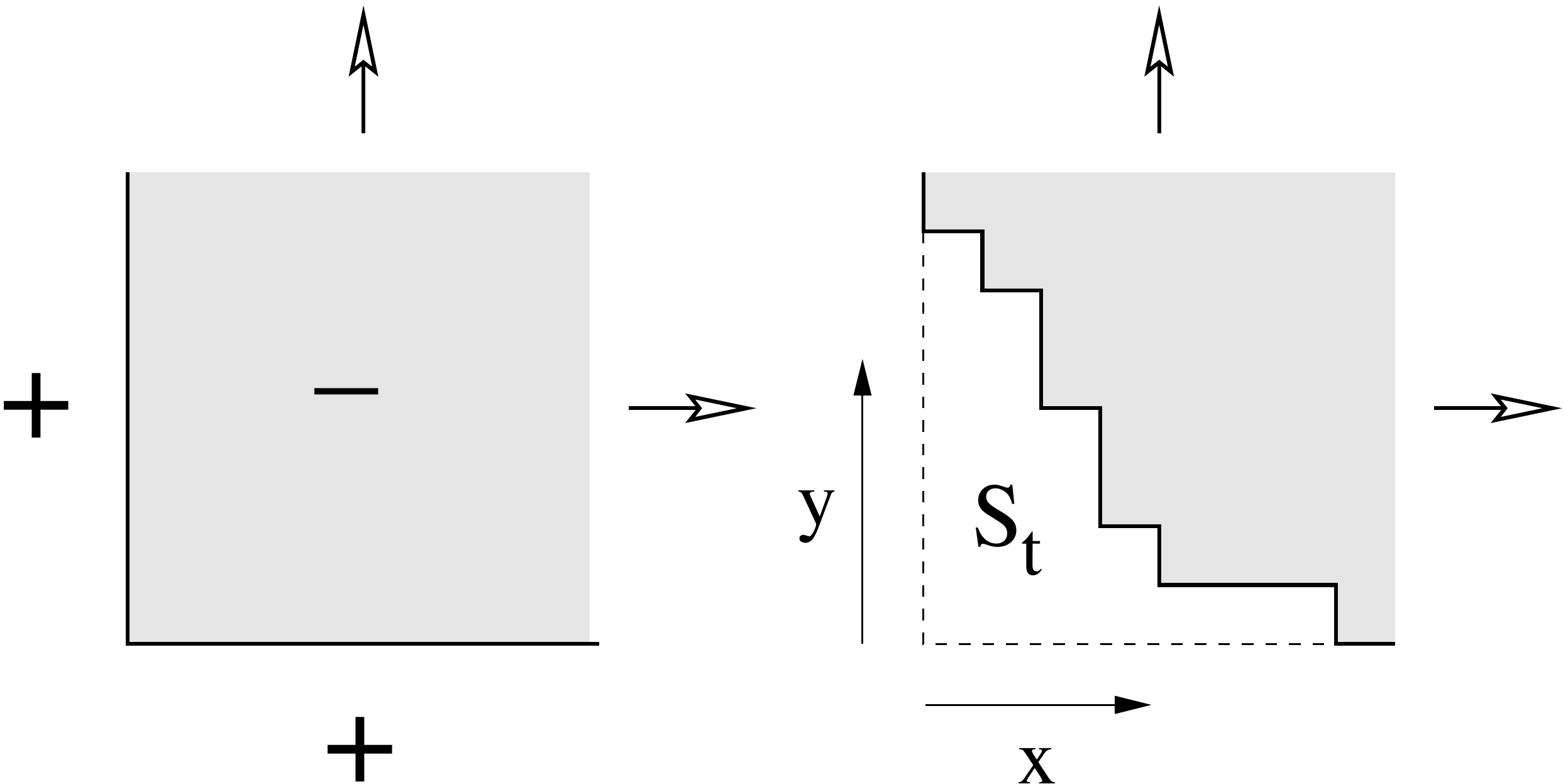}
\caption{The interface is the boundary of a quadrant in the initial state (left); the interface encloses an area $S_t$ at a later time $t$.} 
\label{def-new}
\end{figure}

\begin{figure}
\centering
\includegraphics[scale=0.44]{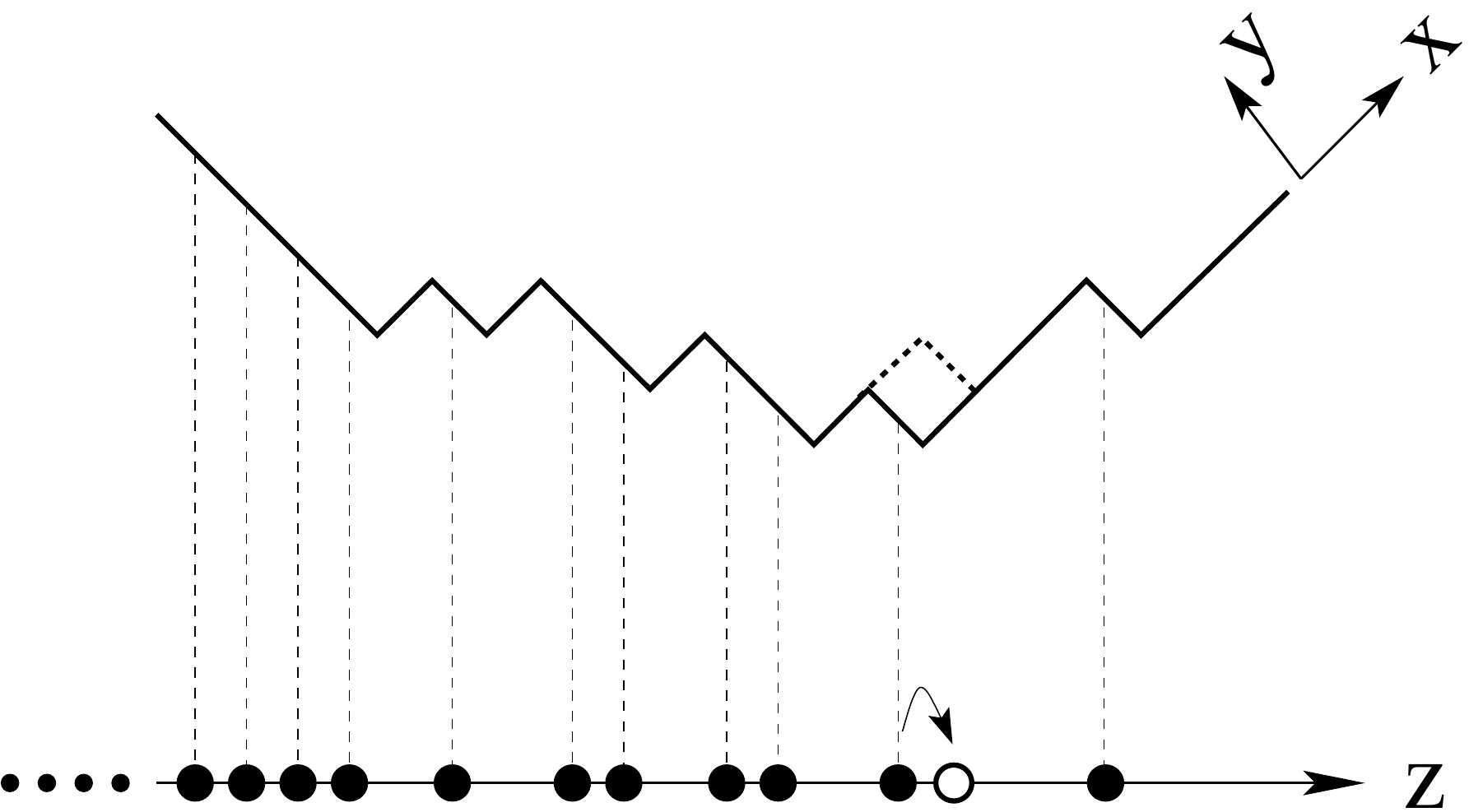}
\caption{An illustrative interface rotated by $\pi/4$ and the corresponding particle configuration. A spin-flip
event is shown together with the corresponding hop of the particle in the lattice gas.} 
\label{csp}
\end{figure}

The zero-temperature dynamics allows spin flips only on interfaces separating ordered phases. This dynamics tends to preserve the integrity of a single interface. Thus, the two-dimensional problems essentially reduce to one-dimensional problems with all the `action' happening along the interface. The above assertions are not manifestly true. Consider a spin surrounded by four similarly aligned spins in an infinite sea of oppositely aligned spins. The interactions of the central spin with its four nearest neighbors must outweigh its interactions with the rest of the misaligned spins to assure that the central spin would not flip. This happens when 
\begin{equation}
\label{JJ}
J_1>\sum_{n\geq 2}nJ_n
\end{equation}
An assumption that the coupling constants rapidly decrease with $n$ presumes the validity of the constraint \eqref{JJ} which obviously guarantees that any spin {\em inside} a droplet of aligned spins cannot flip. The integrity of a single interface is not satisfied for finite interfaces, and even for semi-infinite interfaces bordering regions of finite width (see Sec.~\ref{sec:finger} for more details); this already occurs for the Ising model with zero-temperature dynamics associated with the standard NN interactions. If, however, the size of the interface greatly exceeds the lattice spacing, small (and quickly disappearing) droplets very rarely calve from the interface. In our first example of the melting of a quadrant, the interface manifestly keeps its integrity throughout the evolution.

The dynamics greatly simplifies when we can map the interface onto a one-dimensional lattice gas. Such a mapping is applicable throughout the evolution of a quadrant of minority phase (Fig.~\ref{def-new}). Zero-temperature spin-flip dynamics leads to the melting of the quadrant, and the boundary of the melting quadrant remains an unbounded non-self-intersecting connected interface. The dynamics is stochastic and the interface is also stochastic, yet in the long time limit relative fluctuations vanish and the notion of the limiting shape becomes well-defined.  The goal is to determine the limiting shape of the interface. 

The mapping of the interface dynamics onto a one-dimensional lattice gas is obtained by rotating the interface counter-clockwise by angle $\pi/4$ around the origin and projecting the boundary onto the horizontal line (see Fig.~\ref{csp}). We now identify bonds of the interface with sites on a one-dimensional lattice and we put a particle on a site (leave a site empty) if the corresponding bond on the interface goes along the co-diagonal (diagonal). There could be at most one particle per site (exclusion property). The hopping rates are affected by nearest- and next-nearest neighbors on the one-dimensional lattice in the case where the spin interactions are described by the Hamiltonian \eqref{Ising_Ham} with $k=2$; for an arbitrary $k$, particles separated by distance $\leq k$ from the particle that attempts a hop influence the dynamics. 

More precisely, in the particle mapping for the spin Hamiltonian \eqref{Ising_Ham} with $k=2$, particles can hop to neighboring empty sites (symmetric hopping), and such hops are carried out at random according to the following rules:
\begin{enumerate}
\item If a hop would not increase the number of NN pairs of particles, it is always performed. 
\item If a hop would increase the number of NN pairs of particles, it is never performed.
\end{enumerate}
The above hopping rules directly follow from the zero-temperature spin-flip dynamics compatible with Hamiltonian \eqref{Ising_H} or Hamiltonian \eqref{Ising_Ham} with $k=2$. The major property of the zero-temperature dynamics is that a spin flip is forbidden if it would lead to an increase in energy. There is a bit of arbitrariness in hopping rates, e.g., hops which decrease the number of NN pairs of particles may occur at a different rate than hops which do not change the number of NN pairs of particles. In the long time limit, however, the fraction of hops of the first kind becomes negligible, and therefore it suffices to study the simplest version in which these rates are equal. We shall always set this rate to unity. 

We now illustrate the hopping rules by the early-time dynamics. We start with the unperturbed quadrant, or
\begin{equation}
\label{initial_NNEP}
\ldots \bullet\bullet\bullet\bullet\bullet\circ\circ\circ\circ\circ\ldots
\end{equation}
in the lattice gas representation. The spin at the corner of the quadrant is the only flippable spin. Its flipping is equivalent to the hopping
\begin{equation*}
\ldots \bullet\bullet\bullet\bullet\bullet\circ\circ\circ\circ\circ\ldots \Longrightarrow
\ldots \bullet\bullet\bullet\bullet\circ\bullet\circ\circ\circ\circ\ldots
\end{equation*}
which is the only move compatible with the exclusion property. For the standard Ising model with NN spin-spin interactions, there are three possibilities for the next move:
\begin{equation}
\label{Ising_SEP}
\bullet\bullet\bullet\bullet\circ\bullet\circ\circ\circ\circ  \Longrightarrow
\begin{cases}
\bullet\bullet\bullet\bullet\bullet\circ\circ\circ\circ\circ\\
\bullet\bullet\bullet\circ\bullet\bullet\circ\circ\circ\circ\\
\bullet\bullet\bullet\bullet\circ\circ\bullet\circ\circ\circ
\end{cases}
\end{equation}
which all occur with equal rates. This illustrates the general assertion that the underlying lattice gas is a symmetric simple exclusion process \cite{KD90,barma,book,PK_Ising}. For the model \eqref{Ising_H}, however, there are only two possible moves:
\begin{equation}
\label{Ising_NNN}
\bullet\bullet\bullet\bullet\circ\bullet\circ\circ\circ\circ  \Longrightarrow
\begin{cases}
\bullet\bullet\bullet\circ\bullet\bullet\circ\circ\circ\circ\\
\bullet\bullet\bullet\bullet\circ\circ\bullet\circ\circ\circ
\end{cases}
\end{equation}
Returning back to the original quadrant is impossible since such a flip would raise the energy; according to the lattice gas description, such a move would have created an additional NN pair of particles, and therefore it is never performed. 

Therefore our lattice gas (we shall call it a repulsion process) is governed by the zero-temperature dynamics associated with the Hamiltonian 
\begin{equation}
\label{Ham}
\mathcal{H} = J_2\sum n_i n_{i+1}
\end{equation}
Here $n_i=1$ if site $i$ is occupied, and $n_i=0$ if it is empty. The dynamics is based on the exchange, $n_i \leftrightarrow n_{i+1}$. This exchange is forbidden if it would lead to an increase in the number of NN pairs of particles, and thereby to an increase in energy. 

\section{Melting of the Quadrant} 
\label{sec:quadrant}

Here we consider the Ising ferromagnet with NN and NNN spin interactions.  We apply the above lattice gas representation of the boundary of the corner to the determination of its limiting shape. The first step is to calculate the average density of the lattice gas, which is then used (through the inverse mapping from the lattice gas to the interface) to extract the limiting shape. 

\subsection{Average Density}

The density $n_j(t)$ is a strongly fluctuating quantity.  On the hydrodynamic level we are interested in distances greatly exceeding the lattice spacing. In this situation we can talk about the average density $\rho(z,t)$. This quantity obeys a macroscopic diffusion equation 
\begin{equation}
\label{diff:eq}
\frac{\partial \rho}{\partial t} = \frac{\partial }{\partial z}\!\left[D(\rho)\, \frac{\partial \rho}{\partial z}\right]
\end{equation}
The initial condition \eqref{initial_NNEP} becomes
\begin{equation}
\label{step}
\rho(z, t=0) = 
\begin{cases}
1   & z<0\\
0   & z>0
\end{cases}
\end{equation}
on the macroscopic (hydrodynamic) level.

The diffusion description \eqref{diff:eq} has been justified for numerous lattice gas models \cite{Spohn91,KL99}, but the diffusion coefficient $D(\rho)$ has been computed in very few models, mostly for lattice gases with density-independent diffusion coefficient. For the repulsion process, the diffusion coefficient depends on the density. This already follows from a few exact values:
\begin{equation}
\label{D_exact}
D(\rho) = 
\begin{cases}
1  & \rho=0\\
4  & \rho=\frac{1}{2}\\
1  & \rho=1
\end{cases}
\end{equation}
The predictions of  Eq.~\eqref{D_exact} can be understood as follows. In the small density limit, we can consider one particle in a vacuum. This particle diffuses with $D=1$ (recall that we have set the hopping rates to unity). Similarly when the density is very close to the maximal density $\rho=1$, we can consider a single vacancy in an otherwise fully packed lattice. This vacancy diffuses with $D=1$. Thus $D(\rho=0)=D(\rho=1)=1$. Consider now a small perturbation of the $\rho=\frac{1}{2}$ state. There is just one half-filled equilibrium configuration, $\ldots\bullet\circ\bullet\circ\bullet\circ\bullet\circ\ldots$. A perturbation of the $\rho=\frac{1}{2}$ state is the configuration
\begin{equation*}
\ldots\bullet\circ\bullet\circ\bullet\circ\bullet\bullet\circ\bullet\circ\bullet\circ\bullet\circ\ldots
\end{equation*}
with one doublet. In this configuration both particles forming the doublet can hop. A single hopping event, say to the right, leads to 
\begin{equation}
\circ\bullet\bullet\circ\bullet\circ ~\Longrightarrow~ \circ\bullet\circ\bullet\bullet\circ
\end{equation}
Thus the doublet effectively hops with the same rate 1 as the particles, and since the span of the doublet hops is twice larger (two lattice sites), the diffusion coefficient is 4 times larger. These arguments explain \eqref{D_exact}. 

The diffusion coefficient of an interacting dense lattice gas is usually impossible to determine analytically. Fortunately, for the repulsion process with Hamiltonian \eqref{Ham}, the diffusion coefficient has been recently computed \cite{KR}
\begin{equation}
\label{diffusion}
D(\rho) = 
\begin{cases}
(1-\rho)^{-2}  & 0<\rho<\frac{1}{2}\\
\rho^{-2}       &\frac{1}{2}<\rho<1
\end{cases}
\end{equation}
The derivation \cite{KR} is lengthy. (One must understand the steady states of the repulsion process and then use this understanding to compute the free energy, the compressibility, and employ the Green-Kubo formula  \cite{Spohn91} to determine the diffusion coefficient.) Equation \eqref{diffusion} agrees with \eqref{D_exact} derived above, as well as with the mirror symmetry $D(\rho) = D(1-\rho)$. 

We now return to the boundary-value problem \eqref{diff:eq}--\eqref{step}. The self-similar nature of the problem (viz. the lack of the characteristic spatial scale) tells us that we can seek a solution in a scaling form
\begin{equation}
\label{scaling_diff}
\rho(z,t)=f(\zeta), \quad \zeta=\frac{z}{\sqrt{4t}}
\end{equation} 
The mirror symmetry allows us to limit ourselves to the $\rho<\frac{1}{2}$ region corresponding to the half-line $z>0$ (initially the vacuum state). Inserting \eqref{scaling_diff} into \eqref{diff:eq} and using Eq.~\eqref{diffusion} for the diffusion coefficient, we arrive at the ordinary differential equation for the scaled profile 
\begin{equation}
\label{f:eq}
\frac{d}{d\zeta}\left[(1-f)^{-2}\frac{df}{d\zeta}\right]+2\zeta\,\frac{df}{d\zeta}=0
\end{equation}
while the boundary conditions are
\begin{equation}
\label{f:in}
f(0)=\frac{1}{2}\,, \quad f(\infty) = 0
\end{equation}

\subsection{Limiting Shape}

\begin{figure}[ht]
\centerline{\includegraphics*[width=0.45\textwidth]{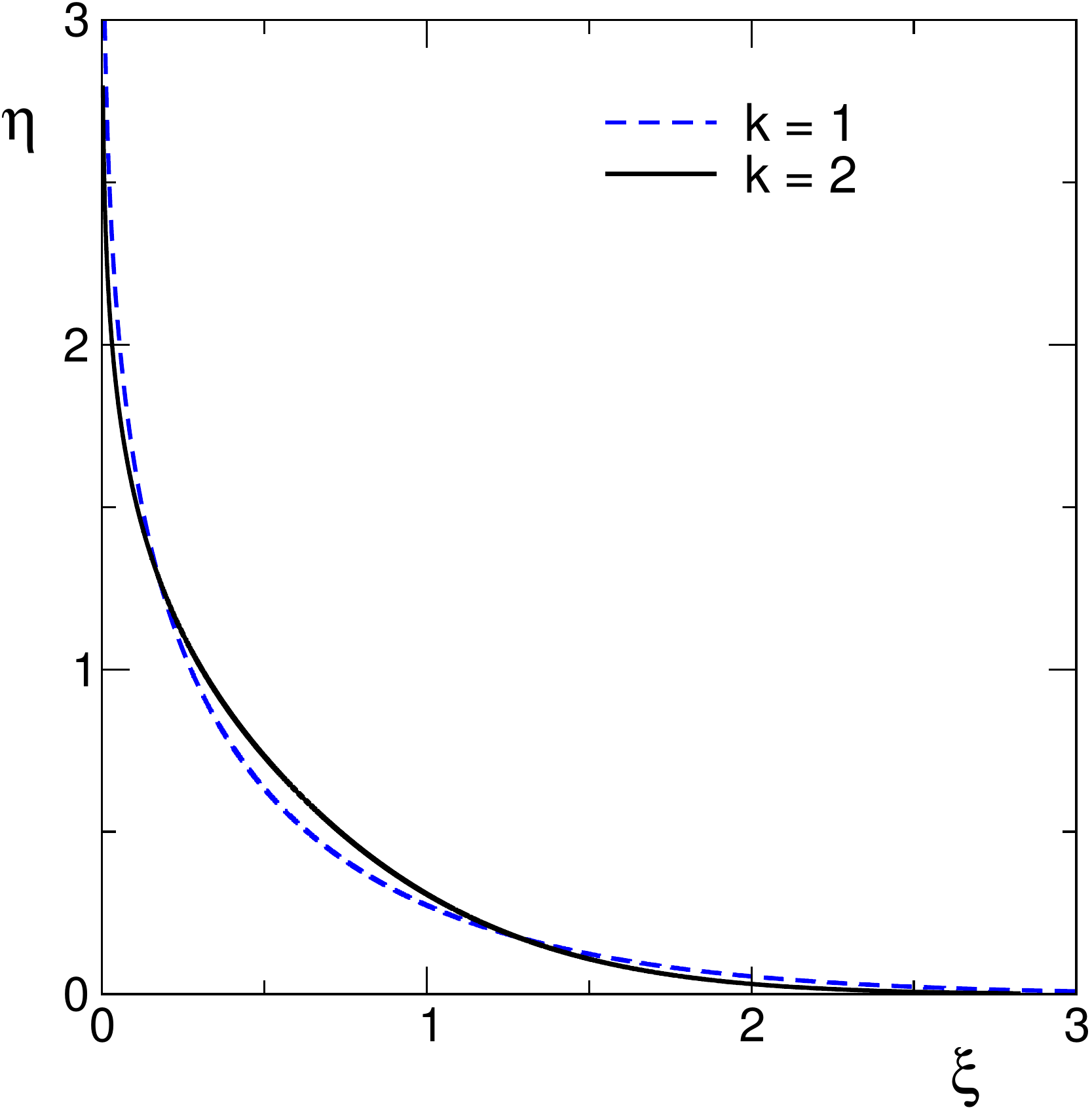}}
\caption{The limiting shape of the melting quadrant corresponding to the Ising model with NNN spin-spin interactions ($k=2$). The limiting shape arising in the standard Ising model ($k=1$) is shown for comparison.  }
  \label{Fig:ising}
\end{figure}

The variables $x$ and $y$ characterizing the interface in the original reference frame can be determined from 
\begin{equation}
\label{yx_inter}
y = \int_{x-y}^\infty dz\,\rho(z,t)
\end{equation}
Writing
\begin{equation}
\label{scaled_coor}
\xi=\frac{x}{\sqrt{4t}}\,,\quad \eta=\frac{y}{\sqrt{4t}}
\end{equation}
we recast \eqref{yx_inter} into 
\begin{equation}
\label{scaled_inter}
\eta=\int_{\xi-\eta}^\infty d\zeta\,f(\zeta)
\end{equation}
which (implicitly) determines the scaled limiting shape of the interface. We solved the boundary-value problem \eqref{f:eq}--\eqref{f:in} numerically. The scaled limiting shape is plotted on Fig.~\ref{Fig:ising}. We also performed numerical simulations of the repulsion process, and we used the discrete version of 
Eq.~\eqref{scaled_inter} to obtain the corresponding limiting shape. We ran 100 realizations of the dynamics, each until time $t=2^{24}$, and we averaged the interface profile over all realizations.  We do not plot simulation results since the discrepancy between the simulated growth process and the numerical solution of the boundary-value problem \eqref{f:eq}--\eqref{f:in} is undetectable by eye.

We now discuss two basic features characterizing the interface --- the intersection of the interface with the diagonal and the area under the interface.

\subsubsection{Distance from the origin}

The diagonal, that is the ray in the $(1,1)$ direction, crosses the interface at 
\begin{equation}
\label{scaled_diag}
\xi_\text{diag} = \eta_\text{diag} = \int_0^\infty d\zeta\,f(\zeta)
\end{equation}
Integrating \eqref{f:eq} from $\zeta=0$ to $\zeta=\infty$, one finds 
\begin{equation}
\label{integral}
\int_0^\infty d\zeta\,f(\zeta) = -2f'(0)
\end{equation}
Combining \eqref{scaled_coor}, \eqref{scaled_diag}, and \eqref{integral}, we arrive at
\begin{equation}
\label{diag}
x_\text{diag} = y_\text{diag} = C\,\sqrt{t}, 
\end{equation}
with
\begin{equation}
\label{diag_ampl}
C = -4f'(0) = 0.8655\ldots 
\end{equation}
More precisely, Eq.~\eqref{diag} is the average position of the intersection of the interface with the diagonal.  The same behavior \eqref{diag} arises for the standard Ising ferromagnet with NN interactions, but in that case the amplitude is $C=\pi^{-1/2}=0.564189583\ldots$. 

\subsubsection{Average area}

The average area is equal to 
\begin{equation}
\label{area}
\langle S_t\rangle =\int_0^\infty y\,dx=4t\int_0^\infty \eta\,d\xi=4t I
\end{equation}
It is convenient to re-write $I$ as a double integral
\begin{equation*}
I =\int \int d\xi\,d\eta=2\int\int_{\xi>\eta>0} d\xi\,d\eta
\end{equation*}
Changing variables, $(\xi,\eta)\to (u=\xi-\eta,\eta)$, and noting that the Jacobian is equal to unity, $\frac{D(u,\eta)}{D(\xi,\eta)}=1$, we get
\begin{equation}
\label{J_integral}
I = 2\int\int_{\xi>\eta>0} d\xi\,d\eta = 2\int_0^\infty \eta\,du
\end{equation}
Plugging \eqref{scaled_inter} into \eqref{J_integral} we get 
\begin{equation}
\label{I_integral}
I = 2\int_0^\infty du \int_u^\infty d\zeta\,f(\zeta)=2\int_0^\infty d\zeta\,\zeta f(\zeta)
\end{equation}
We now multiply Eq.~\eqref{f:eq} by $\zeta$ and integrate from $\zeta=0$ to $\zeta=\infty$. Using integration by parts we obtain
\begin{eqnarray*}
2\int_0^\infty d\zeta\,\zeta f(\zeta) &=& -\frac{1}{2} \int_0^\infty d\zeta\,(1-f)^{-2}\frac{df}{d\zeta}\\
&=& \frac{1}{2} \int_0^{1/2} \frac{df}{(1-f)^2} = \frac{1}{2}
\end{eqnarray*}
Thus $I=\frac{1}{2}$ and therefore the average area is 
\begin{equation}
\label{area_2}
\langle S_t\rangle = 2t
\end{equation}
Interestingly, the average area in the case of the standard Ising ferromagnet, $\langle S_t\rangle = t$, is twice smaller. 

In Fig.~\ref{Fig:ising} we show the limiting shapes for the standard Ising ferromagnet and for the Ising ferromagnet with NNN spin interactions. We re-scale the plots in such way that in the $(\xi,\eta)$ variables both areas are equal to unity: $\int_0^\infty \eta\,d\xi=1$. Hence instead of \eqref{scaled_coor} the scaling variables are $(\xi,\eta)=t^{-1/2}(x,y)$ in the case of the standard Ising ferromagnet and $(\xi,\eta)=(2t)^{-1/2}(x,y)$ for the Ising ferromagnet with NNN spin interactions.

\subsection{Inner and Outer Corners}

Initially there is one inner corner and no outer corners (left part of Fig.~\ref{def-new}).  Generally we denote by $N_+(t)$ [resp. $N_-(t)$] the average total number of inner [resp. outer] corners. The conservation law
\begin{equation}
\label{excess}
N_+(t) - N_-(t) =1
\end{equation}
has an elementary topological origin and it is always valid in two dimensions. The average numbers of inner and outer corners grow as $\sqrt{t}$. This in conjunction with the conservation law \eqref{excess} leads to conjectural behaviors
\begin{subequations}
\begin{align}
\label{N+_Ising_2d}
N_+ &= A_2 \sqrt{t} +  B_2+1 + \ldots \\
\label{N-_Ising_2d}
N_- &= A_2 \sqrt{t} +  B_2  + \ldots
\end{align}
\end{subequations}

The leading asymptotic in \eqref{N+_Ising_2d}--\eqref{N-_Ising_2d} is 
\begin{equation}
\label{A_C}
A_2=2C
\end{equation}
with $C$ given by \eqref{diag_ampl}. Indeed, let's start at the diagonal point and go to $x\to \infty$. The interface will make downward steps of length one; more precisely, almost all steps will have length one in the long time limit. The reason is easy to see in the realm of the lattice gas representation. The gas is in local equilibrium, and equilibrium at a density $\rho<\frac{1}{2}$ has a simple structure \cite{KR}: Adjacent particles are always separated by at least one vacancy. Therefore the total number of inner corners below the diagonal point is equal to $y_\text{diag}$ in the leading order. Similarly, the total number of inner corners above the diagonal point is equal to $x_\text{diag}$. Hence $N_+ =x_\text{diag}+y_\text{diag}$ in the leading order, which implies \eqref{A_C}. 

We do not have a rigorous justification of \eqref{N+_Ising_2d}--\eqref{N-_Ising_2d}, although our simulations support these expansions. It would be interesting to determine $B_2$ analytically. 

\section{Influence of a Magnetic Field} 
\label{sec:quadrant_field} 

Here we consider the quadrant in an external magnetic field. We assume that the magnetic field favors the majority phase (otherwise the quadrant does not evolve). For the Ising model with NN and NNN spin interactions, the Hamiltonian reads
\begin{equation}
\label{Ham_H}
\mathcal{H} = - J_1 \sum_{|{\bf i}-{\bf j}|=1} s_{\bf i} s_{\bf j} - J_2 \sum_{|{\bf i}-{\bf j}|=2} s_{\bf i} s_{\bf j}
- h \sum_{{\bf i}} s_{\bf i}
\end{equation}
with positive magnetic field, $h>0$, if the majority phase is the plus phase (as in Fig.~\ref{def-new}). At zero temperature, the strength of the magnetic field is irrelevant as long as it is smaller than a certain threshold so that only spins at the inner corners can (in principle) flip. In this situation, in every move one minority spin flips. In the lattice gas representation, the dynamics is the biased version of the dynamics studied previously, viz. the hopping rules are identical up to the additional constraint --- only hopping to the right is allowed. 

The hydrodynamic description of the biased repulsion process is based on the continuity equation
\begin{equation}
\label{continuity}
\frac{ \partial\rho}{\partial t} +\frac{\partial J}{\partial  z} =0
\end{equation}
The average current $J(\rho)$ has been computed \cite{KR}
\begin{equation}
\label{current}
J(\rho) = 
\begin{cases}
\frac{\rho(1-2\rho)}{1-\rho}  & 0<\rho<\frac{1}{2}\\
\frac{(1-\rho)(2\rho-1)}{\rho}  &\frac{1}{2}<\rho<1
\end{cases}
\end{equation}
The current is maximal when $\rho^*=\frac{1}{\sqrt{2}}$ and  $\rho_*=1-\frac{1}{\sqrt{2}}$, with
$J(\rho^*)= J(\rho_*) = J_{\rm max} =3-2\sqrt{2}$. The relation between the maxima, $\rho_* + \rho^*=1$, reflects the more general mirror symmetry, $J(\rho) = J(1-\rho)$. Note that for the standard Ising model in a magnetic field the corresponding lattice gas is an asymmetric simple exclusion process, as was first discovered by Rost \cite{Rost}. 

The derivation of \eqref{current} is simple since the steady states of the biased repulsion process are known (they are the same as in the symmetric case) \cite{KR}. The expression \eqref{current} for the current, sometimes only in the high-density regime $\rho>\frac{1}{2}$, appears in a number of models \cite{B91,Krug91,AS00,BM09,GKR}.

We solve Eq.~\eqref{continuity} by noting that the density admits a scaling form
\begin{equation}
\label{scaling_hydro}
\rho(z,t)=R(Z), \quad Z=\frac{z}{t}
\end{equation}
Plugging this scaling form into \eqref{continuity} and using Eq.~\eqref{current} for the current we obtain a rarefaction wave
\begin{equation}
\label{R_Wave}
R(Z) = 
\begin{cases}
1                     & ~~~~~~Z <-1\\
(2+Z)^{-1/2}    &\, -1<  Z <0\\
1-(2-Z)^{-1/2}  &~~~0 < Z <1 \\
0          &\, ~~~~~~~~Z >1
\end{cases}
\end{equation}
when the initial condition is given by \eqref{step}. 

The density profile of the rarefaction wave \eqref{R_Wave} is discontinuous at $x=0$, with densities $\rho(-0)=\rho^*=\frac{1}{\sqrt{2}}$ and $\rho(+0)=\rho_*=1-\frac{1}{\sqrt{2}}$, precisely the values where the current attains its maximum. The average total number of particles that penetrates into the region $x>0$ is therefore $N(t) = J_{\rm max} t$, where $J_{\rm max}=3-2\sqrt{2}$

\begin{figure}[ht]
\centerline{\includegraphics*[width=0.45\textwidth]{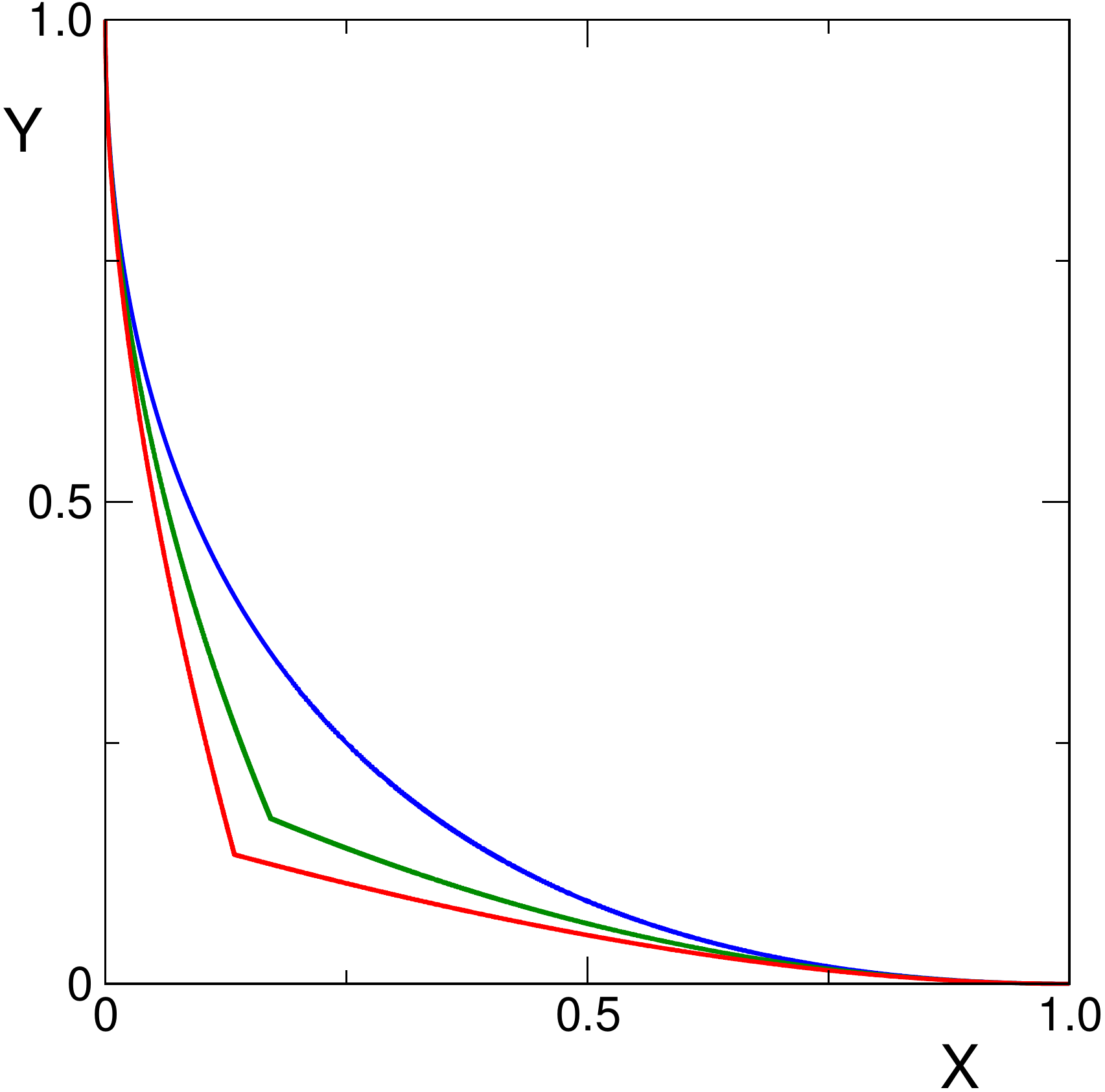}}
\caption{Limiting shapes of the melting quadrant in a magnetic field. Upper curve: The limiting shape $\sqrt{X}+\sqrt{Y}=1$ corresponding to the standard Ising model, see \cite{Rost}. 
Middle curve: The limiting shape \eqref{LS_field} corresponding to $k=2$. 
Lower curve: The limiting shape Eq.~\eqref{LS_field_3} corresponding to $k=3$.}
\label{Fig:growth}
\end{figure}

The interface is determined from \eqref{yx_inter}, which becomes 
\begin{equation}
\label{YX_inter}
Y=\int_{X-Y}^\infty dZ\, R(Z)
\end{equation}
in the scaled variables
\begin{equation}
\label{scaled_XY}
X=\frac{x}{t}\,,\quad Y=\frac{y}{t}
\end{equation}
Combining \eqref{YX_inter} with \eqref{R_Wave}, we find that the limiting shape of the interface (see Fig.~\ref{Fig:growth}) is given by 
\begin{equation}
\label{LS_field}
Y(X) = 
\begin{cases}
1- 2\sqrt{X}                   & 0\leq X\leq J_{\rm max}\\
(1-2X+X^2)/4                &J_{\rm max} \leq X \leq 1
\end{cases}
\end{equation}

Alternatively, one can directly derive, see Appendix~\ref{Ap:EE}, an equation describing the evolution of the interface in the long time limit (when we can ignore fluctuations):
\begin{equation}
\label{EE}
y_t = 
\begin{cases}
-y_x(1+y_x)      & 0> y_x > -1\\
 1+(y_x)^{-1}    & -1>y_x
\end{cases}
\end{equation}
where $y_t=\frac{\partial y}{\partial t}$ and $y_x=\frac{\partial y}{\partial x}$. The structure of \eqref{EE} suggests to seek a solution in the scaling form $Y=Y(X)$. Plugging this form into \eqref{EE}, we obtain a differential equation for $Y(X)$ whose solution is given by Eq.~\eqref{LS_field}.

\section{Ising Finger}
\label{sec:finger}

\begin{figure}
\includegraphics[scale=0.55]{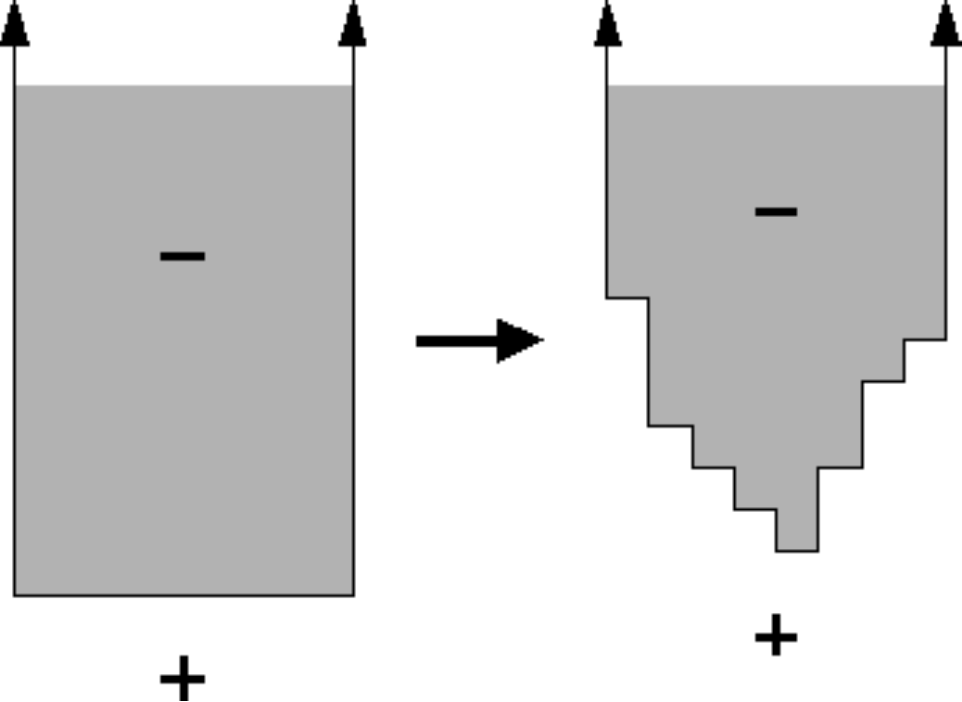}
\caption{Schematic illustration of the evolution of a semi-infinite strip (a rectangular finger of the minority minus phase surrounded by the majority plus phase). On the right-hand side, the flip of the lowest minority spin, the tip spin in this example, is an irreversible process that causes the minimum height of the finger to advance by one.} 
\label{finger_ill}
\end{figure}

We now turn to the finger geometry (Fig.~\ref{finger_ill}). We assume that the minority phase initially occupies the semi-infinite region: $y>0$ and $|x|<L$.  In the interesting long time regime, more precisely when $t\gg L^2$, the two corners of the initial finger interact and the finger relaxes to a limiting shape moving along the axis of the finger (in the $y>0$ direction) with some constant velocity.  In a reference frame moving with the finger, the interface $y(x)$ is stationary. We shall determine this stationary limiting shape by employing a continuum description which is valid when the width of the finger greatly exceeds the lattice spacing: $L\gg 1$.

\begin{figure}
\centering
\includegraphics[scale=0.64]{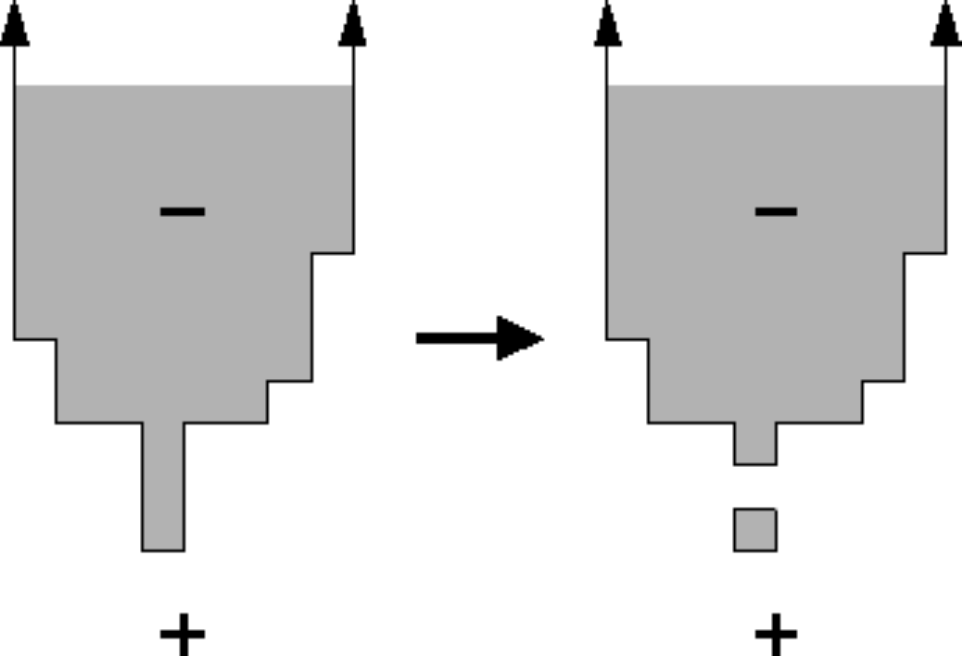}
\caption{An illustration of a rare emergence of a short-living tiny closed interface during the evolution of the finger.} 
\label{shedding}
\end{figure}

During the evolution of the finger one rarely observes a fission of tiny domains of minority phase from the finger. Apart from pathological initial conditions, e.g. those which contain `tendrils' of width one (i.e. equal to the lattice spacing) or strips of width one, or rare separations of tiny drops (Fig.~\ref{shedding}), such break ups play a significant role only for narrow strips, $L=\mathcal{O}(1)$. In the interesting situation of $L\gg 1$ we can ignore fission events and fluctuations and focus on the deterministic limiting shape.

Due to symmetry we can limit ourselves to the region $0<x<L$ and $y>0$. The governing equation for the interface shape $y(x,t)$ is [see Appendix~\ref{Ap:EE}]
\begin{subequations}
\begin{align}
\label{2d_small}
y_t &= y_{xx}\,, \qquad 0<y_x<1\\
\label{2d_large}
y_t &= \frac{y_{xx}}{y_x^2}\,, \qquad  1<y_x
\end{align}
\end{subequations}
Consider an upward moving reference system in which the interface is stationary. In the bottom part of the interface, $0<x<L'$ with yet unknown $L'$, we should use \eqref{2d_small}. Therefore
\begin{subequations}
\begin{eqnarray}
\label{y_bottom}
v=y_{xx}\,, \quad y(0)= y_x(0) = 0, \quad y_x(L')=1
\end{eqnarray}
In the top part, $L'<x<L$, we employ \eqref{2d_large}, so we have
\begin{eqnarray}
\label{y_top}
v=\frac{y_{xx}}{y_x^2}\,, \quad y_x(L')=1, \quad y(L)=y_x(L)=\infty
\end{eqnarray}
\end{subequations}
We must solve \eqref{y_bottom} and \eqref{y_top}, match the solutions, and self-consistently determine the intermediate point $L'$ and the velocity $v$ of the finger. 

Transforming the variables
\begin{equation}
X = \frac{x}{L}\,, \quad Y = \frac{y}{L}\,, \quad V = vL, \quad \ell  = \frac{L'}{L}
\end{equation}
and writing $F=\frac{dY}{dX}$ we recast \eqref{y_bottom}--\eqref{y_top} into
\begin{subequations}
\begin{align}
\label{F_bottom}
V&=\frac{dF}{dX}\,, \quad F(0)=0, \quad F(\ell)=1\\
\label{F_top}
V&=F^{-2}\,\frac{dF}{dX}\,, \quad F(\ell)=1, \quad F(1)=\infty
\end{align}
\end{subequations}

Integrating \eqref{F_bottom}--\eqref{F_top}  we obtain
\begin{subequations}
\begin{align}
\label{F_bottom_sol}
F& =VX, \quad V\ell =1 \\
\label{F_top_sol}
F& = \frac{1}{1-V(X-\ell)}
\end{align}
\end{subequations}
Combining \eqref{F_top_sol} with $F(1)=\infty$ and $V\ell =1$ we get
\begin{equation}
\label{VL}
V = 2\,, \quad \ell = \frac{1}{2}
\end{equation}
Thus the velocity of the finger is 
\begin{equation}
\label{finger-velocity}
v = \frac{2}{L}
\end{equation}
For the Ising ferromagnet with NN spin-spin interactions, the velocity $v=1/L$ (see \cite{PK_Ising}) is twice slower. This two-fold acceleration is analogous to the twice faster growth of the area, Eq.~\eqref{area_2}, in the corner problem. 

\begin{figure}[ht]
\centerline{\includegraphics*[width=0.45\textwidth]{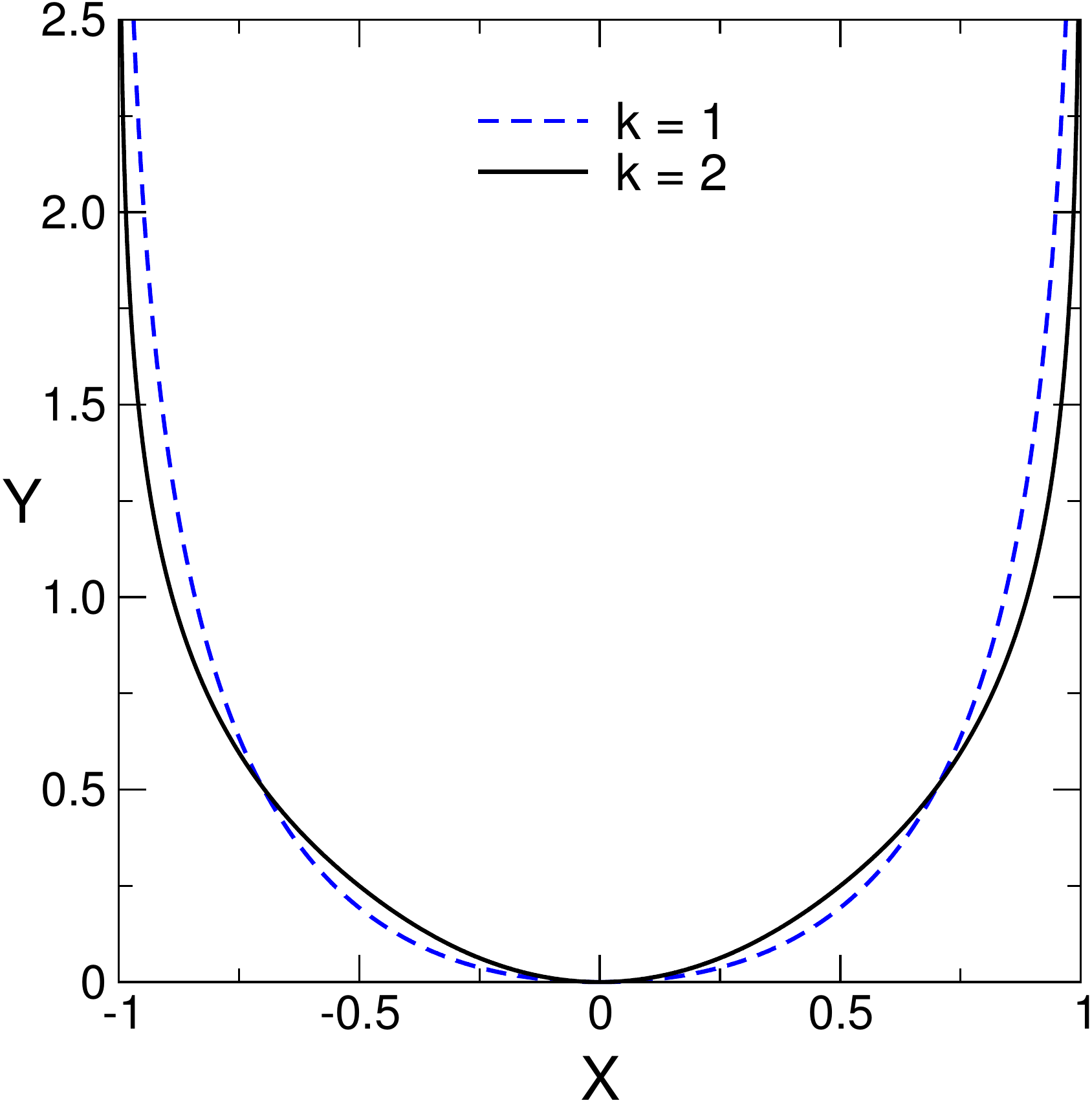}}
\caption{The limiting shape, Eq.~\eqref{LS_finger}, of the finger. The limiting shape $Y=-|X|-\ln(1-|X|)$ (see \cite{PK_Ising}) of the finger in the case of the standard Ising dynamics is shown for comparison (dotted blue curve).}
  \label{Fig:Fin}
\end{figure}

Using Eqs.~\eqref{F_bottom_sol}--\eqref{F_top_sol} and \eqref{VL}, we integrate $F=\frac{dY}{dX}$ and find the shape of the finger (see Fig.~\ref{Fig:Fin})
\begin{equation}
\label{LS_finger}
Y = 
\begin{cases}
X^2                                                      & |X|\leq \frac{1}{2}\\
\frac{1}{4} - \frac{1}{2}\,\ln(2-2|X|)       &\frac{1}{2} \leq |X| \leq 1
\end{cases}
\end{equation}

\section{Long-Ranged Interactions}
\label{sec:range}

Here we study Ising ferromagnets with longer interaction range. Specifically, we consider the class of Hamiltonians \eqref{Ising_Ham} with $k\geq 3$. When  $k=3$, the corresponding lattice gas has NN and NNN repulsive interactions 
\begin{equation}
\label{Ham_3}
H_3 =  J_2\sum_{i=-\infty}^\infty n_i n_{i+1} + J_3\sum_{i=-\infty}^\infty n_i n_{i+2} 
\end{equation}
In the general case,
\begin{equation}
\label{Ham_k}
H_k =  J_2\sum_{i=-\infty}^\infty n_i n_{i+1} + \ldots+J_k\sum_{i=-\infty}^\infty n_i n_{i+k-1} 
\end{equation}
Note that the interaction span for the repulsion process \eqref{Ham_k} is equal to $k-1$, while in the underlying Ising ferromagnet the interaction span between the spins is $k$.

The correspondence between the Hamiltonians \eqref{Ising_Ham} of the two-dimensional Ising ferromagnets and the Hamiltonians  \eqref{Ham}, \eqref{Ham_3} and \eqref{Ham_k} of one-dimensional lattice gases is not straightforward, e.g. the Hamiltonians \eqref{Ising_Ham} involve the interactions in the bulk (away from the interface) which are omitted. What is important is that the non-conservative spin-flip dynamics caused by \eqref{Ising_Ham} is the same, in terms of the interface, as the conservative hopping dynamics implied by the Hamiltonian \eqref{Ham} for $k=2$,  \eqref{Ham_3} for $k=3$, and \eqref{Ham_k}  for general $k$. 

Narratively, the dynamics corresponding to \eqref{Ham_3} is the following:  
\begin{enumerate}
\item If a hop would decrease the number of NN pairs of particles, it is always performed. 
\item If a hop would not change the number of NN pairs of particles, and it would not increase the number of NNN pairs of particles, it is always performed. 
\item Otherwise a hop is never performed.
\end{enumerate}
This dynamics arises independently of the coupling constants as long as the inequality $J_2>J_3$ is obeyed. Generally the positive coupling constants in \eqref{Ham_k} are assumed to obey the set of constraints
\begin{equation}
\label{JJJ}
J_a>J_{a+1}+\ldots+J_k, \quad a=2,\ldots, k -1
\end{equation}
These constraints simplify the analysis since they allow us to treat interactions in a lexicographic order. 

The analysis of the repulsive process with Hamiltonian \eqref{Ham_3} is rather cumbersome, yet the results are neat and simple. In the totally asymmetric case (only hopping to the right is allowed), the current $J(\rho)$ is given by \cite{KR}
\begin{equation}
\label{current_3}
J(\rho) = 
\begin{cases}
\frac{\rho(1-3\rho)}{1-2\rho}       & 0<\rho<\frac{1}{3}\\
\frac{(1-2\rho)(3\rho-1)}{\rho}     & \frac{1}{3}<\rho<\frac{1}{2}\\
\frac{(2\rho-1)(2-3\rho)}{1-\rho}  & \frac{1}{2}<\rho<\frac{2}{3}\\
\frac{(1-\rho)(3\rho-2)}{2\rho-1}  &\frac{2}{3}<\rho<1
\end{cases}
\end{equation}
The diffusion coefficient reads \cite{KR}
\begin{equation}
\label{diffusion_3}
D(\rho) = 
\begin{cases}
(1-2\rho)^{-2}       & 0<\rho<\frac{1}{3}\\
\rho^{-2}              & \frac{1}{3}<\rho<\frac{1}{2}\\
(1-\rho)^{-2}        & \frac{1}{2}<\rho<\frac{2}{3}\\
(2\rho-1)^{-2}      &\frac{2}{3}<\rho<1
\end{cases}
\end{equation}
in the symmetric version. 

Generally for the repulsive process with Hamiltonian \eqref{Ham_k}, the current is given by \cite{KR}
\begin{equation}
\label{current:GRP}
J(\rho) =
\begin{cases}
\frac{\rho(1-k\rho)}{1-(k-1)\rho}                 &  0<\rho<\frac{1}{k}\\
\frac{[(j + 1)\rho-1] (1-j\rho)}{\rho}               &  \frac{1}{j+1}<\rho<\frac{1}{j}\\
\frac{[j-(j + 1)\rho](j\rho-j+1)}{1-\rho}         &  \frac{j-1}{j}<\rho<\frac{j}{j+1}\\
\frac{(1-\rho)(k\rho-k+1)}{(k-1)\rho-k+2}   &  \frac{k-1}{k}<\rho<1
\end{cases}
\end{equation}
Here $j=2,3,\ldots, k-1$. The diffusion coefficient characterizing the models with symmetric hopping is \cite{KR}
\begin{equation}
\label{diff:GRP}
D(\rho) = 
\begin{cases}
[1-(k-1)\rho]^{-2}         & 0<\rho<\frac{1}{k}\\
\rho^{-2}                     & \frac{1}{k}<\rho<\frac{1}{2}\\
(1-\rho)^{-2}               & \frac{1}{2}<\rho<\frac{k-1}{k}\\
[(k-1)\rho-k+2]^{-2}    &\frac{k-1}{k}<\rho<1
\end{cases}
\end{equation}

In Secs.~\ref{sec:quadrant}--\ref{sec:finger}, we computed the limiting shapes corresponding to the Ising ferromagnet with NNN interactions ($k=2$). We now briefly describe limiting shapes arising in situations when $k>2$. 

\subsection{Melting of the Quadrant} 

We use the same notations as in Sec.~\ref{sec:quadrant}. The scaled density profile satisfies 
\begin{subequations}
\begin{align}
\label{f_head}
&(f^{-2}f')' + 2\zeta f' = 0, \quad 0<\xi<\ell\\
\label{f_tail}
&([1-(k-1)f]^{-2} f')'+ 2\zeta f' = 0, \quad \ell<\xi<\infty
\end{align}
\end{subequations}
where $\ell$ is the scaled position on the one-dimensional lattice where the density crosses $1/k$ and the diffusion coefficient \eqref{diff:GRP} changes its dependence on the density. The boundary conditions are
\begin{equation}
\label{BC:k}
\quad f(0)=\tfrac{1}{2}, \quad f(\ell)=\tfrac{1}{k}, \quad f(\infty)=0
\end{equation}
The same analysis as in Sec.~\ref{sec:quadrant} shows that the diagonal point on the interface is given by \eqref{diag} with $C=-4f'(0)$. The value of the amplitude depends on $k$ and it can be determined by numerically solving Eqs.~\eqref{f_head}--\eqref{f_tail} subject to the boundary conditions \eqref{BC:k}.

It is again possible to compute the average area avoiding a numerical solution of the boundary-value  problem, Eqs.~\eqref{f_head}--\eqref{f_tail} and \eqref{BC:k}, and even without finding $\ell$. Equations \eqref{area} and \eqref{I_integral} are still applicable. To compute $I$ we multiply Eqs.~\eqref{f_head}--\eqref{f_tail} by $\zeta$ and integrate the former from $\zeta=0$ to $\zeta=\ell$ and the latter from $\zeta=\ell$ to $\zeta=\infty$. Using integration by parts we find 
\begin{equation*}
4\int_0^\infty d\zeta\,\zeta f(\zeta) = \int_0^{1/k} \frac{df}{[1-(k-1)f]^2} + \int_{1/k}^{1/2} \frac{df}{f^2} = k-1
\end{equation*}
Thus $I=(k-1)/2$ and hence the average melted area [see Eq.~\eqref{area}] is given by
\begin{equation}
\label{area_k}
\langle S_t\rangle = 2(k-1)t
\end{equation}

\subsection{Quadrant in a Magnetic Field}

Here we consider the quadrant in a magnetic field favoring the majority phase.  Consider first the situation when $k=3$. Inserting the scaling form \eqref{scaling_hydro} into Eq.~\eqref{continuity} and using Eq.~\eqref{current_3} for the current, we obtain a rather simple expression for the scaled density
\begin{equation}
\label{R_Wave_3}
R(Z) = 
\begin{cases}
1                                                                 & ~~~~~~Z <-1\\
\frac{1}{2}\left[1+(3+2Z)^{-1/2}\right]          &\, -1<  Z <0\\
\frac{1}{2}\left[1-(3-2Z)^{-1/2}\right]            &~~~0 < Z <1 \\
0          &\, ~~~~~~~~Z >1
\end{cases}
\end{equation}
For $Z<0$, we have $\rho>\tfrac{1}{2}\big(1+1/\sqrt{3}\big)>\tfrac{2}{3}$, while for $Z>0$ we have
$\rho<\tfrac{1}{2}\big(1-1/\sqrt{3}\big)<\tfrac{1}{3}$. Therefore only the behaviors of the current in the regions $0<\rho<\frac{1}{3}$ and $\frac{2}{3}<\rho<1$ matter. 

Generally, the current vanishes at $2k-1$ points, but again only the behavior of the current in the regions between the first two and last two zeros matter, viz. the behavior of $J(\rho)$ when $0<\rho<\frac{1}{k}$ and $\frac{k-1}{k}<\rho<1$. The generalization of \eqref{R_Wave} and \eqref{R_Wave_3} reads 
\begin{equation}
\label{R_Wave_k}
R(Z) = 
\begin{cases}
1                                                                 & ~~~~~~Z <-1\\
\frac{k-2+[k+(k-1)Z]^{-1/2}}{k-1}                 &\, -1<  Z <0\\
\frac{1-[k-(k-1)Z]^{-1/2}}{k-1}                      &~~~0 < Z <1 \\
0          &\, ~~~~~~~~Z >1
\end{cases}
\end{equation}

Using Eq.~\eqref{YX_inter} we find that the diagonal point on the interface, $X_\text{diag}=Y_\text{diag}\equiv X_k$, is given by
\begin{equation}
X_k=\int_0^1 dZ\,\frac{1-[k-(k-1)Z]^{-1/2}}{k-1}=\frac{1}{(\sqrt{k}+1)^2}
\end{equation}
Combining Eq.~\eqref{YX_inter} and \eqref{R_Wave_k} we find the limiting shape $Y=Y(X)$ in the range $X_k<X<1$:
\begin{equation}
\label{LS_k}
Y = \frac{k-(k-2)X-2\sqrt{k-1-(k-2)X}}{(k-2)^2}
\end{equation}
In the region $0<X<X_k$ the limiting shape is found by using the mirror symmetry: 
\begin{equation}
\label{LS_k_mirror}
X = \frac{k-(k-2)Y-2\sqrt{k-1-(k-2)Y}}{(k-2)^2}
\end{equation}
In particular, for $k=3$ the limiting shape is given by
\begin{equation}
\label{LS_field_3}
\begin{split}
Y &= 3-X-2\sqrt{2-X}\,, \quad (\sqrt{3}+1)^{-2}<X<1\\
X &= 3-Y-2\sqrt{2-Y}\,, \quad 0<X<(\sqrt{3}+1)^{-2}
\end{split}
\end{equation}
This limiting shape is plotted on Fig.~\ref{Fig:growth}. 

\subsection{Ising Fingers}

In the reference system in which the finger is stationary, the governing equations for the interface are
\begin{equation}
v= y_{xx}\times 
\begin{cases}
[1-(k-2)y_x]^{-2}            & 0<y_x<\frac{1}{k-1}\\
y_x^{-2}                         & \frac{1}{k-1}< y_x <1\\
1                                    &1<y_x<k-1 \\
[y_x-k+2]^{-2}                & k-1<y_x<\infty
\end{cases}
\end{equation}

Using the same treatment and notation as in Sec.~\ref{sec:finger}, one finds the velocity of the finger 
\begin{equation}
\label{finger-V}
v = \frac{V}{L}\,, \qquad V = 2(k-1)
\end{equation}
The finger is composed of four pieces on the right from the tip, $0<X<\ell_1, ~\ell_1<X<\ell_2, ~\ell_2<X<\ell_3$ and $\ell_3<X<1$, supplemented by four analogous pieces on the left ($-1<X<0$). The separation points are  
\begin{equation*}
\ell_1= \frac{1}{V}\,, \quad \ell_2=\frac{1}{2}\,, \quad \ell_3=1-\ell_1
\end{equation*}
The limiting shape can be expressed through elementary functions. The bottom part ($0<|X|<\ell_1$) of the finger 
\begin{equation}
\label{L1}
Y = \frac{|X|}{k-2}-\frac{1}{V(k-2)^2}\,\ln[1+V(k-2)|X|]
\end{equation}
is close to a parabola, $Y\simeq \frac{1}{2}VX^2$, near the tip. The other parts are
\begin{equation}
\label{L2}
Y=C_1-V^{-1}\ln(k-V|X|)
\end{equation}
for $\ell_1<|X|<\ell_2$, 
\begin{equation}
\label{L3}
Y=C_2 + |X|+\tfrac{1}{2}V(|X|-\ell_2)^2
\end{equation}
for $\ell_2<|X|<\ell_3$, and
 \begin{equation}
\label{L4}
Y = C_3 + (k-2)|X| -V^{-1}\ln(1-|X|)
\end{equation}
for $\ell_3<|X|<1$. The constants $C_1, C_2, C_3$ are found by matching the shapes at common separation points. For instance, matching  \eqref{L1} and  \eqref{L2} at $|X|=\ell_1$, one finds
\begin{equation*}
C_1 = \frac{1}{V(k-2)} - \frac{1}{V(k-2)^2}\,\ln(k-1)+\frac{1}{V}\,\ln(k-1)
\end{equation*}

\section{Concluding Remarks}
\label{concl}

We investigated a zero-temperature spin-flip dynamics of the Ising model on the square lattice with ferromagnetic spin-spin interactions spanning beyond nearest neighbors. Specifically, we studied the evolution of a single interface separating two ordered phases. We showed that the mapping of an unbounded interface onto a lattice gas leads to an interesting family of lattice gases with repulsive interactions. These repulsion processes are tractable \cite{KR}, and this allowed us to determine the limiting shape of the interface in the situation where initially the interface is either the boundary of the quadrant, or the boundary of a semi-infinite stripe with width greatly exceeding the lattice spacing. We found that the limiting shape of interfaces in the Ising model depend on the interaction range. Thus we obtained an infinite family of different limiting shapes for the melting quadrant (with and without magnetic field) and for the Ising finger.

An important and feasible extension of our work is to consider the shrinking of a droplet of one phase in a sea of the opposite phase. This droplet, evolving according to a zero-temperature spin-flip dynamics, will eventually disappear. In the intermediate regime when the droplet is still very large compared to the lattice spacing, but already very small compared to the initial size, the droplet has an essentially deterministic shape (fluctuations are negligible in comparison with the characteristic size of the droplet) and the initial condition no longer matters. In the case of the classical Ising model, this limiting shape has been analytically determined \cite{PK_Ising}; see also \cite{Alex} and the proof of the convergence to the limiting shape \cite{Toni}. It appears possible to generalize these results to include long-ranged interactions. 

Another obvious extension is to three dimensions. This is a much more challenging domain, e.g., little is known about the fate of three-dimensional Ising ferromagnets quenched to zero temperature \cite{SKR,OKR}. Limiting shapes are also unknown, and even evolution equations governing the motion of interfaces haven't been established. The mapping of interfaces onto two-dimensional lattice gases is possible, but it has not yet led to success. In the simplest case of the melting octant in the presence of a magnetic field, the governing evolution equation for the interface has been proposed in Ref.~\cite{JPSK}. It would be interesting to generalize this conjectural evolution equation to the situations with long-ranged interactions. 

\bigskip\noindent
We are grateful to S. Redner for discussions and collaboration at an earlier stage of this work. The research of J.O. is supported by NSF Grant No. DMR-1205797.

\appendix
\section{Equations Governing the Evolution of Interfaces}
\label{Ap:EE}

Consider first the simpler case of the evolution in a magnetic field. In the lattice gas framework, the governing equation is the continuity equation \eqref{continuity}.  The current is given by Eq.~\eqref{current} in our lattice gas, or by $J=\rho(1-\rho)$ for the asymmetric exclusion process which arises for the standard Ising model with only NN spin interactions.  The strategy is therefore to solve \eqref{continuity} subject to the proper initial condition, like Eq.~\eqref{step} for the corner case, and then to use \eqref{yx_inter} to extract the limiting shape. Here we establish a closed equation describing the evolution of the interface. To deduce such an equation for the interface $y=y(x,t)$, we re-write \eqref{yx_inter} as  
\begin{equation}
\label{interface_eq}
y(x,t)=\int_{x-y(x,t)}^\infty dz\,\rho(z,t)
\end{equation}
Taking the time derivative of \eqref{interface_eq}, we obtain
\begin{equation}
\label{interface_t}
y_t = \rho y_t  + \int_{x-y}^\infty dz\,\rho_t = \rho y_t  + J(\rho)
\end{equation}
where on the second step we have integrated $\rho_t = -J_z$.  Differentiating \eqref{interface_eq} with respect to $x$, we obtain
\begin{equation}
\label{interface_x}
y_x = \rho (y_x-1)
\end{equation}
Hence we can express the lattice gas density through the slope of the interface, $\rho=y_x/(y_x-1)$. Using this result we recast Eq.~\eqref{interface_t}  into a closed evolutionary equation 
\begin{equation}
\label{interface_field}
y_t = (1-y_x)\,J\!\left(\frac{y_x}{y_x-1}\right)
\end{equation}
When the current is given by Eq.~\eqref{current}, the general equation \eqref{interface_field} reduces to Eq.~\eqref{EE}. 

Without a magnetic field, the governing equation in the lattice gas framework is the diffusion equation \eqref{diff:eq}. Taking the time derivative of \eqref{interface_eq}, we obtain
\begin{equation}
\label{interface_tx:long}
y_t =\rho y_t  - D(\rho)\,\frac{\partial \rho}{\partial z}\Big|_{z=x-y}
\end{equation}
Using $dz=dx(1-y_x)$ and $1-\rho=(1-y_x)^{-1}$, we recast Eq.~\eqref{interface_tx:long} into
\begin{equation}
\label{interface_tx}
y_t = - D(\rho) \rho_x
\end{equation}
Relation \eqref{interface_x} remains valid and this allows one to reduce \eqref{interface_tx} into a closed evolutionary equation 
\begin{equation}
\label{interface_main}
y_t = D\!\left(\frac{y_x}{y_x-1}\right)\,\frac{y_{xx}}{(1-y_x)^2}
\end{equation}
When the diffusion coefficient is given by \eqref{diffusion}, the general equation \eqref{interface_main} reduces to \eqref{2d_small}--\eqref{2d_large}.


\begin{thebibliography}{99}
\bibitem{long}
    For long-ranged spin-spin interactions $U$ decaying as a power-law, $U\propto r^{-\sigma}$, 
    with the separation $r$ between spins, the equilibrium critical behavior changes when 
    the exponent $\sigma$ becomes smaller than a certain threshold. We shall always assume 
    that interactions vanish beyond a certain threshold, or that they fall down very rapidly, 
    e.g. exponentially. This setting provides the 
    most stringent test of the universality of the standard model with only nearest-neighbor interactions.
\end{thebibliography}
\end{document}